# Une interprétation du spin en mécanique relativiste

A relativistic interpretation of the spin

**Stefan Catheline**

**Affiliation : LabTAU, INSERM, Centre Léon Bérard, Université Lyon 1, Univ Lyon, F-69003, LYON, France**

ORCID : 0000-0002-5524-4120        stefan.catheline@inserm.fr

## Résumé

L'objet d'étude de ce manuscrit est le spin. Le point de départ est donc l'expérience de Stern et Gerlach dont les résultats n'ont pu être décrits de façon cohérente que dans le cadre de la mécanique quantique. A l'instar d'un précédent article portant sur la rotation solide, le point de vue adopté dans ce manuscrit est celui de la mécanique relativiste. L'étude des effets de perspective de l'horizon de la rotation solide relativiste fait apparaitre une invariance vis-à-vis de l'angle d'observation satisfaisant pleinement à la notion de spin.

## Introduction

La mécanique quantique est intimement liée au concept de spin introduit dans l'expérience de Stern et Gerlach en 1922 [1]. En effet, aucune description cohérente de physique classique n'a pu rendre compte de la quantification du moment cinétique des particules, faisant par conséquent du spin un être purement quantique. Le spin justifie donc à lui seul les concepts abstraits et contre-intuitifs de la mécanique quantique pour lesquels « on n'éprouve jamais le sentiment confortable » qu'ils soient « naturels » selon Richard Feynman [2]. L'objectif de cette section n'est pas d'en donner une explication de mécanique classique mais relativiste. L'électron est la particule chargée la plus légère. Elle se définit entièrement par les données de sa charge élémentaire, de sa masse et de son spin. C'est cette particule porteuse de spin qui, dans les atomes d'argent des expériences originelles de Stern et Gerlach pilote leur trajectoire. Nous empruntons parfois dans le texte le raccourci commode qui consiste à mentionner l'électron plutôt que l'atome d'argent. Depuis 1820 et les expériences d'Arago-Ampère, le magnétisme est décrit comme conséquence de boucles de courant. Nous adoptons ici le point de vue classique des pionniers de la mécanique quantique et admettons comme modèle de l'électron une sphère en rotation pourvu d'une charge e en rotation circulaire à la vitesse de la lumière c sur une orbite de rayon $r_e$ et donc d'intensité $I = \frac{ec}{\pi r_e}$. La rotation de la charge e à l'origine du magnétisme de spin est donc subordonnée à la notion de rotation solide relativiste introduite dans une publication antérieure [3]. Cette connection spin-rotation désuète pour certains ou hérétique pour d'autres, est le point de départ de ce manuscrit qu'il reste bien entendu à démontrer. Ce sera l'objet de discussion d'une future publication. Pour l'heure, concentrons notre attention sur la rotation solide en relativité qui a fait l'objet de multiples articles [4,5,6,7]. L'originalité de l'approche présentée dans ce manuscrit tient à son objet d'étude qui porte exclusivement sur le comportement de la limite de la rotation solide, c'est-à-dire de son horizon. Nous nous attacherons particulièrement à montrer que ses propriétés géométriques permettent de revisiter l'expérience de Stern et Gerlach.

## Rappel de rotation solide relativiste

La rotation solide relativiste ayant déjà fait l'objet d'une publication antérieure [3], nous rappelons ici rapidement les propriétés utiles à notre démonstration. Il s'agit d'une rotation solide composée d'une rotation angulaire $\vec{\omega}$, d'une vitesse $\vec{v}$, et un rayon vecteur $\vec{r}_0$, obéissant à l'équation $\vec{v} = \vec{\omega} \wedge \vec{r}_0$, figure 1. Son originalité tient au fait que le chemin imposé par l'accélération de Coriolis à la lumière émise depuis le centre O le long de la direction radiale $\vec{r}$ subit une distorsion angulaire θ par rapport au vecteur vitesse rendant impossible les valeurs supérieures à celle de la lumière c. Ce chemin radial est un simple cercle dit « d'espace ». L'arc de cercle d'espace reliant le centre O à la position du satellite s'obtient par l'angle de rotation Ω et le demi-rayon de rotation $\frac{\vec{R}}{2}$ selon : $\vec{r} = \vec{\Omega} \wedge \frac{\vec{R}}{2}$. A travers la formulation de l'effet Doppler transverse, le cercle d'espace est décrit depuis 1963 [8]. Il a ensuite été confirmé par la méthode du tenseur métrique d'espace emprunté à la relativité [9]. Lorsque la vitesse v du satellite est faible devant celle de la lumière, on retrouve le cas classique de la rotation solide. Lorsqu'au contraire la vitesse est proche de celle de la lumière, le satellite atteint son rayon maximum R, c'est l'horizon de la rotation solide relativiste. Enfin la rotation solide relativiste satisfait pleinement les transformations de Lorentz de l'espace et du temps. La distorsion angulaire θ entre la direction de la vitesse et la tangente au cercle d'espace modifie les perceptions d'espace et de temps depuis le référentiel inertiel central O. Les paramètres relativistes sont liés à la géométrie de la figure 2 selon les équations $\cos\theta = \frac{r_0}{R} = \frac{v}{c}$ d'une part et $\sin\theta = \sqrt{1 - \frac{v^2}{c^2}}$ d'autre part. La contraction de Lorentz d'une longueur $L_0$ alignée à la vitesse est perçue depuis un référentiel inertiel placé au centre O comme $L_\varphi = L_0 \sin\theta$. Contrairement à la rotation classique, la rotation solide relativiste n'implique qu'une région finie de l'espace. Elle possède donc par construction une vitesse maximum : la première des constantes universelles c découle donc de la nature géométrique finie de la rotation solide relativiste. Bien que majeures, ces distinctions ne justifient pas pour autant la nécessité d'employer la rotation solide relativiste plutôt que la relativité restreinte dont la compatibilité est bien établie [10,11]. Cependant la première permet l'interprétation de géométrie Euclidienne simple et directe du spin comme expression de son cercle d'horizon.

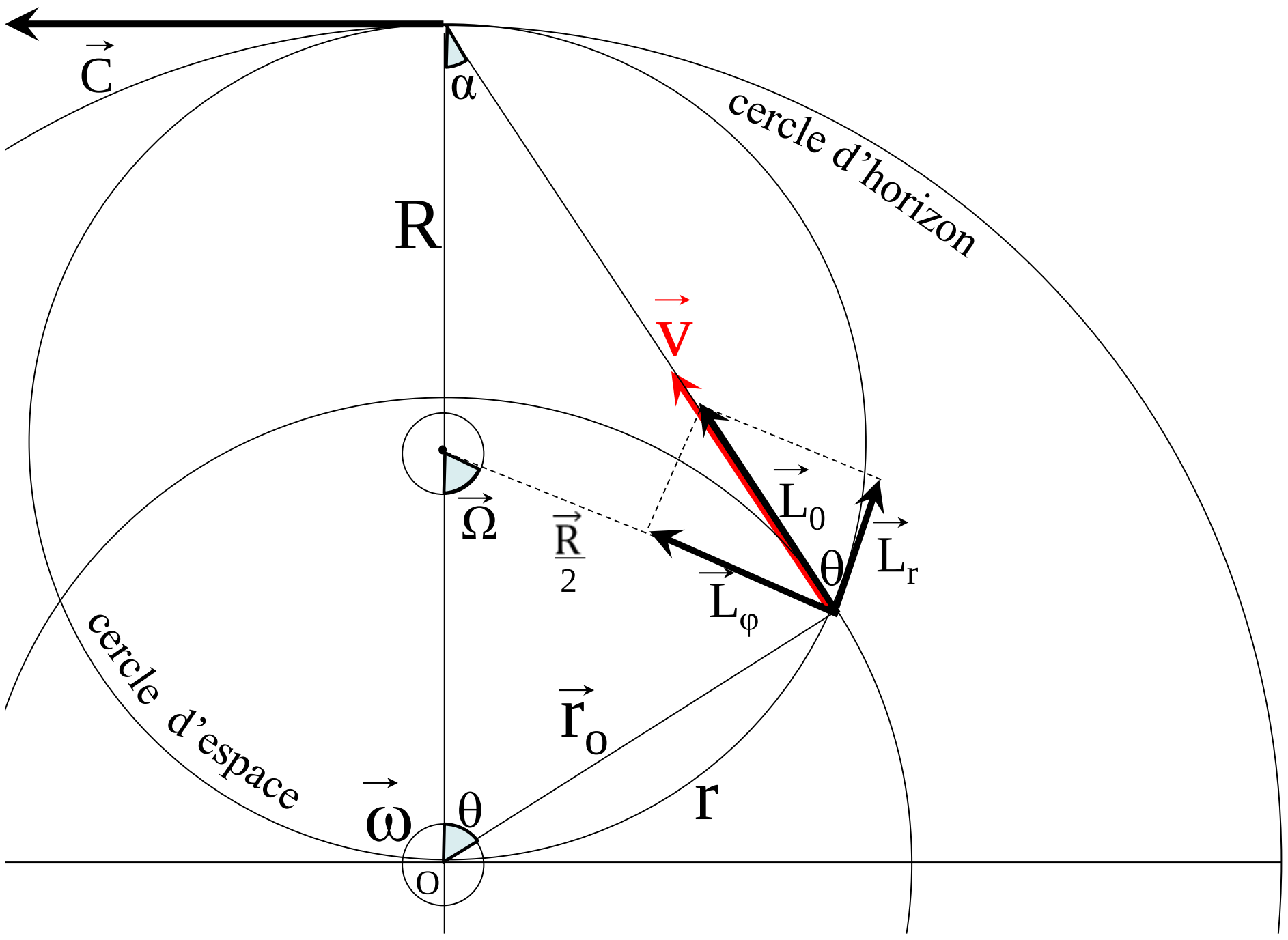

*Figure 1 : Rotation solide relativiste. Sa dimension est finie et bordée par un cercle d'horizon. Les propriétés de cet horizon sont à l'origine du spin.*

## Echec de l’interprétation classique des expériences de Stern et Gerlach

Prenons le cas d’un faisceau d’atomes d’argent neutre traversant le montage de Stern et Gerlach, figure 2. Concentrons notre étude à la seule propriété de spin que lui confère un électron. Inspiré d’un modèle classique d’une sphère de densité de charge en rotation, l’électron se compose d’une charge e voyageant à la vitesse de la lumière c sur une orbite située à un rayon électronique $r_e$. Cette boucle de courant d’intensité $I = \frac{ec}{\pi r_e}$ subit une force lorsqu’elle est immergée dans un gradient de champ magnétique et est déviée de son mouvement rectiligne uniforme. Rappelons au passage la pertinence de l’approche classique du calcul du spin par une boucle de courant. En effet, le moment cinétique sur l’horizon de la rotation relativiste est la constante de Planck, $m_e r_e c = \bar{h}$. Le produit de l’intensité par la surface donne donc, $IS = e r_e c = 2\frac{e\bar{h}}{2m_e}$. Enfin, compte tenu du rapport gyromagnétique de l’électron on obtient bien comme moment magnétique de spin $\mu_B = \frac{e\bar{h}}{2m_e}$ appelé magnéton de Bohr-Procopiu. En vertu de la force d’induction classique d’un circuit plongé dans un champ magnétique et décrite par l’équation suivante : $\vec{F} = -\overrightarrow{\text{grad}}\left(I\,\vec{B}.\vec{S}\right) = IS\cos\theta\,\overrightarrow{\text{grad}}\,B$, cette boucle va migrer selon son orientation vers les régions de champ magnétique plus ou moins dense en cherchant à optimiser les lignes de flux à travers sa surface. Toutes les directions d’orientation θ du disque de rotation de la boucle de courant par rapport au champ magnétique étant équiprobables avant la traversée du montage, la surface $S_B$ offerte au flux du gradient de champ magnétique suit une distribution continue de type $S_B = S\cos\theta$. Cet effet de perspective est le même pour un circuit circulaire que pour un circuit carré montré sur la figure 3 à des fins pédagogiques et pour lequel un côté de longueur $L_0$ surligné en trait plein devient pour un observateur latéral $L = L_0\cos\theta$. Lors de la traversée, une distribution de forces conforme à l’effet de perspective aléatoire perçu par un observateur posté parallèlement aux lignes de champ magnétique s’applique donc à chaque circuit. Il en découle que la déviation classique attendue d’un faisceau d’électrons par un gradient de champ magnétique donne une distribution spatiale continue. L’expérience montre une séparation en deux du faisceau incident ce qui infirme le raisonnement précédent. Nombre de physiciens l’affirment : toute explication mécanique à la description du spin est vaine [12], toute approche intuitive est vouée à l’échec.

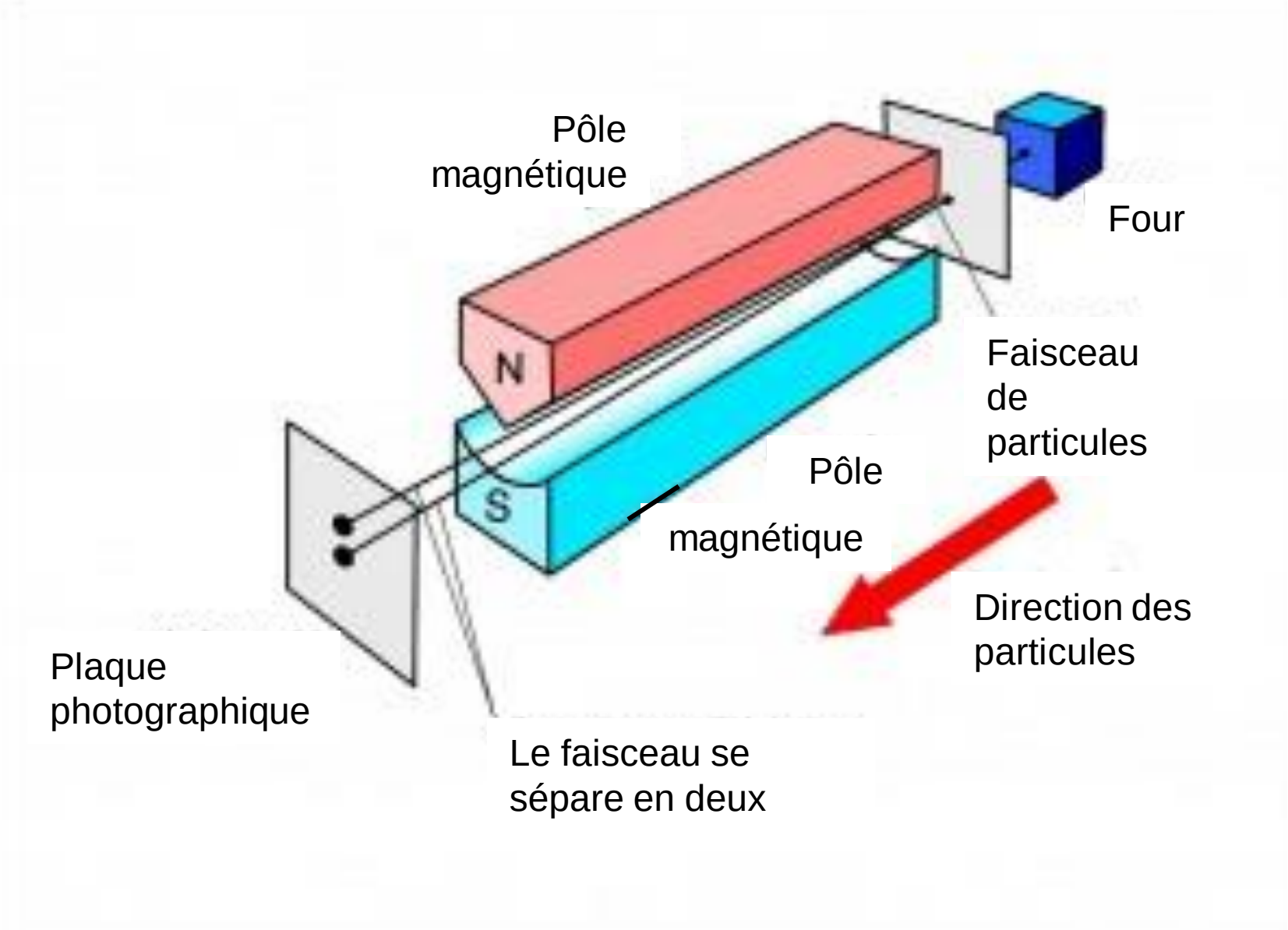

*Figure 2 : Montage expérimental de Stern et Gerlach. La division en deux du faisceau incident de particules, visible sur l'écran de sortie ne trouve pas d'explication cohérente en physique classique.*

## Emergence du spin

Reprenons notre raisonnement sur le flux de champ magnétique traversant un circuit et examinons plus attentivement les effets relativistes sur cette surface. Connu en peinture depuis la renaissance italienne sous le nom de règle du point de fuite, l'effet de perspective est une homothétie qui diminue la taille apparente des objets en raison inverse de l'éloignement à l'observateur. Dans le cadre de la rotation solide relativiste, figure 2, un satellite de longueur propre L se situe à la distance $r_0 = R\cos\theta$, sa taille apparente, en vertu de l'énoncé précédent sur la règle du point de fuite, est donc proportionnelle à $\frac{L}{R\cos\theta}$. Lorsqu'une mesure de longueur est faite, la règle métrique utilisée par l'observateur placé en O dans un référentiel inertiel, possède une longueur propre $L_0$ lorsque sa vitesse est minimum, $v_{min} \approx 0$, $\theta \approx \frac{\pi}{2}$, figure 2. La règle est placée à la même distance que le satellite de sorte qu'elle subit exactement la même réduction de perspective $\propto \frac{L_0}{R\cos\theta}$ et la mesure classique de la longueur L est inchangée par l'éloignement. Cependant, dans la rotation solide relativiste, une distorsion apparait car l'éloignement est synonyme d'accroissement du champ de vitesse. Pour l'observateur placé en O, l'alignement de la mesure est la direction perpendiculaire au cercle d'espace, ce dernier matérialisant la direction radiale qu'emprunte la lumière. Dans le cas classique de vitesses faibles, l'alignement de la mesure peut se confondre avec la direction de la vitesse mais, en toute rigueur, une distorsion angulaire apparait. Cette distorsion angulaire θ diminue par projection de $L_0$ la longueur de la règle de mesure selon l'égalité $L_\varphi = L_0 \sin\theta$ : c'est la contraction des longueurs. Quand cette règle contractée est utilisée pour mesurer le périmètre P d'un cercle que l'on divise par son rayon, on trouve l'inégalité non-euclidienne suivante : $\frac{P}{r_0} > 2\pi$. Plus quantitativement, on retrouve géométriquement l'expression relativiste [13] $\frac{P}{r_0} = \frac{2\pi}{\sin\theta}$. Lorsque la vitesse avoisine la limite c, ce périmètre d'horizon est maximum et définit le contour géométrique du spin de l'électron.

Examinons à présent l'effet de perspective classique sur un tel contour en commençant par un circuit carré. Le segment du côté de longueur $L_0$ surligné en gris sur la vue de dessus de la figure 3a est parcouru par une particule animée de la vitesse c. Lors d'une rotation d'angle θ du circuit selon son axe de symétrie vertical, ce segment apparait vu de côté contracté par la projection de la rotation $L = L_0 \cos\theta$. Par le même effet de perspective, la projection de la vitesse de la particule sur cette portion donne $v = c \cos\theta$. C'est aussi la définition de l'angle de distorsion relativiste [3]. Cette coïncidence de θ comme angle de rotation de perspective et angle de distorsion relativiste de vitesse, est cruciale dans la démonstration car cela permet de représenter sur le plan Euclidien de la figure 3 l'effet de la perspective sur l'horizon des circuits carré et circulaires. C'est précisément l'objectif de cet article. Commençons par la vue de face du circuit carré noir. La perspective contracte ses côtés horizontaux seulement et devient le rectangle surligné d'un liseré rouge. Les longueurs des cotés parcourues à la vitesse apparente v ne peuvent être, par définition, les frontières de l'horizon. Il faut pour rétablir la vitesse maximum c propre à l'horizon appliquer la loi de la rotation solide à travers une homothétie de raison $\frac{1}{\cos\theta} = \frac{c}{v}$ depuis le centre du circuit O. On obtient après dilatation le nouveau rectangle en ligne pointillée rouge. C'est l'horizon recherché. Les mêmes étapes appliquées au circuit circulaire noir de la figure 3b donne l'ellipse à liseré rouge par perspective puis l'ellipse en ligne pointillée après l'application de l'homothétie afin de retrouver l'horizon. Les deux cas de figure aboutissent à la constatation suivante : la surface $S = LL' = \frac{v}{c} L_0 \frac{c}{v} L_0 = L_0^2$ du rectangle horizon est aussi celle du circuit carré et la surface $S = \pi ab = \pi \frac{v}{c} r_e \frac{c}{v} r_e = \pi r_e^2$ de l'ellipse horizon est aussi celle du circuit circulaire. Autrement dit les effets de perspectives changent la forme de l'horizon sans en changer la surface. La conséquence immédiate est la suivante : la surface offerte au ligne de flux magnétique ne variant pas avec la perspective, le module de la force magnétique qui s'applique au circuit est la même quel que soit son orientation. Seul le signe de la force dépend de l'orientation des lignes de courant par rapport au champ magnétique. Un seul module avec deux signes possibles : c'est exactement ce qui est attendu du spin.

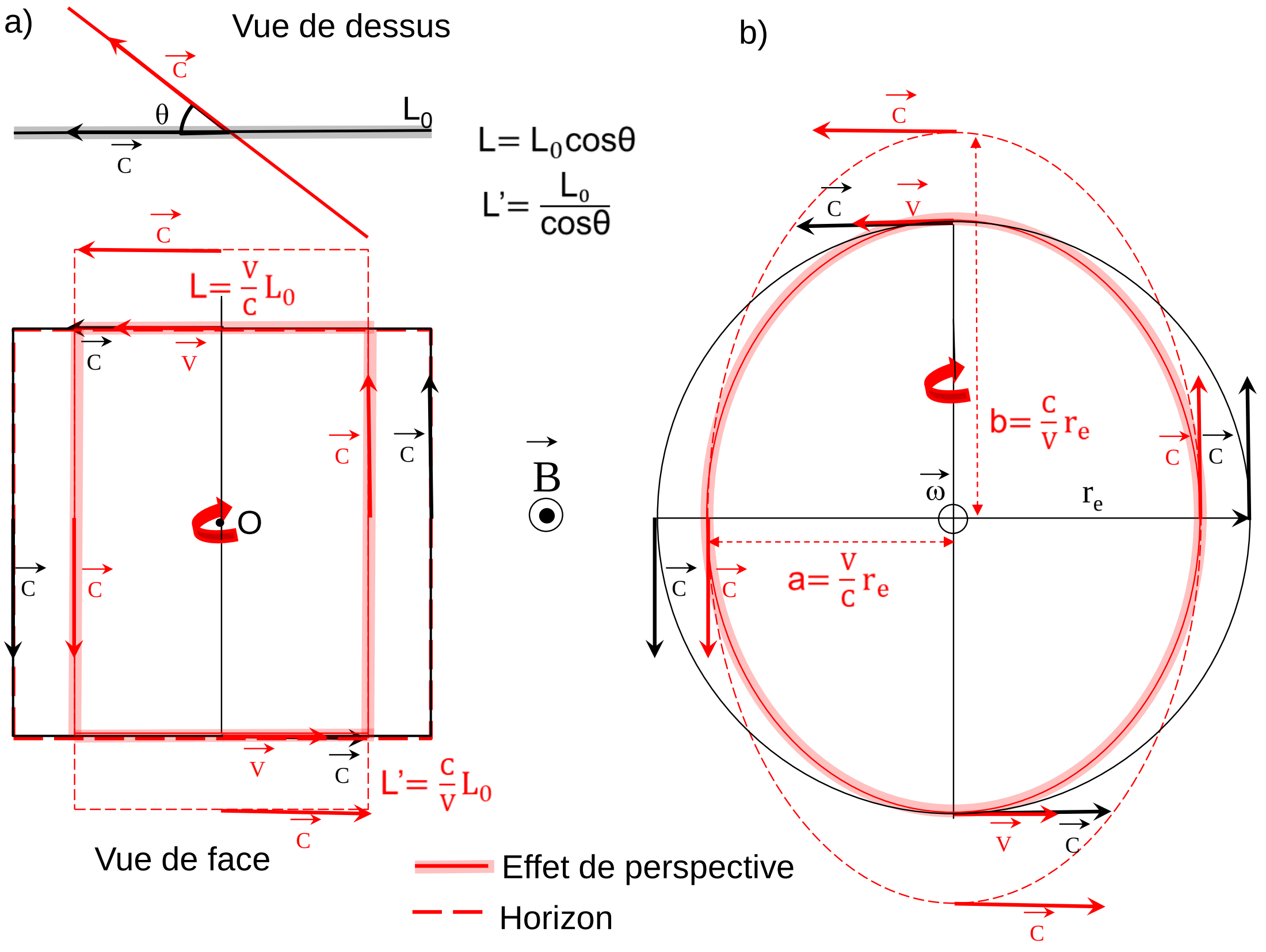

*Figure 3 : a) Vue de dessus et vue de face de la rotation d'angle $\theta$ d'un circuit carré utilisé à des fins pédagogiques. Sur la vue de face, les effets de perspective classique transforment le carré en rectangle avec le liseré rouge. L'horizon de ce circuit est représenté par le rectangle pointillé. b) Effet de perspective d'un circuit circulaire. La rotation d'un cercle selon son diamètre vertical donne une ellipse avec liseré rouge. Sur son horizon parcouru par des particules circulant à la vitesse c, l'ellipse en pointillé est dilatée de la quantité $\frac{1}{\cos\theta} = \frac{c}{v}$.*

En résumé, les effets de perspective angulaire d'une rotation solide relativiste pour un observateur qui se déplace autour de son centre O modifie la forme de l'horizon qui, d'un cercle, se transforme comme dans le cas classique en ellipse mais plus longue et dont l'aire est invariante : l**a surface circonscrite par l'horizon d'une rotation solide relativiste est la même depuis toutes les directions d'observations**. Il s'ensuit que la force d'induction qui découle du flux du champ magnétique stationnaire à travers cette surface invariante est constante.

Elle ne dépend que de son orientation parallèle ou antiparallèle au gradient du champ magnétique donnant lieu à une séparation du faisceau en deux dans l'expérience de Stern et Gerlach. C'est l'effet *spin*. Le secret du spin réside donc tout entier dans la propriété d'invariance de la surface d'une rotation solide relativiste vis-à-vis de l'angle de vue. Cette idée fondamentale d'invariance angulaire de la surface délimitée par l'horizon d'une rotation solide relativiste ramène donc dans le giron de la relativité, l'un des concepts fondateurs de la mécanique quantique.

## Test du spin relativiste

S'il est vrai que la surface balayée par l'horizon de la rotation solide relativiste ne varie pas en fonction de l'angle d'observation, il n'en va pas de même pour sa forme. Cette dernière propriété peut être testée de la façon suivante : lorsqu'elle est observée selon sa tranche, la rotation circulaire devient une ellipse qui s'allonge jusqu'à atteindre et dépasser la taille de la région dans laquelle règne le gradient du champ magnétique. Dès lors, la surface offerte au ligne de flux diminue ainsi que la force magnétique qui en découle. La déviation des électrons diminue pour tendre vers zéro. Même si la majorité des électrons se rassemblent sur deux points de l'écran à la sortie du montage expérimental de Stern et Gerlach, il existe quelques électrons qui se présentent sur la tranche à l'entrée du montage et donc dévient classiquement entre ces deux points selon une distribution uniforme. Plus quantitativement, l'angle de vue à partir duquel le grand axe de l'ellipse s'étire sur toute la région du champ magnétique $\Delta l$ est donné par la relation $\frac{r_e}{\cos\theta} = \Delta l$ soit $\theta_{lim} = \arccos\frac{r_e}{\Delta l}$. En faisant l'hypothèse d'une distribution angulaire homogène des spins du faisceau incident et en prenant une région du gradient du champ magnétique de largeur millimétrique, $\Delta l = 10^{-3}$mm, les particules vues sur la tranche, $\theta > \theta_{lim}$, représentent une proportion de l'ensemble des particules d'environ $1 - \frac{\theta_{lim}}{\frac{\pi}{2}} \approx 10^{-10}$. La mesure de la répartition spatiale de ces particules dans un montage de Stern et Gerlach en fonction du nombre de particules incidentes ou des caractéristiques géométriques du champ magnétique représenterait un test de la thèse défendue dans ce manuscrit.

## Conclusion

Pour revisiter le spin par la mécanique relativiste, nous avons adopté le point de vue originel d'une charge en rotation à la vitesse de la lumière sur une orbite circulaire appelé horizon. La démonstration que contient ce manuscrit porte exclusivement sur la surface délimitée par cet horizon de rotation solide relativiste en fonction de l'angle d'observation. Il est apparu qu'en dépit du changement de forme de la ligne d'horizon, sa surface était invariante. Par conséquent, dans une expérience de Stern et Gerlach, le flux des lignes de champ magnétique traversant une telle surface est constant. Cela explique que le module de la force d'induction s'appliquant au circuit ne dépende plus de son orientation. Il s'agit là de la propriété de spin qui puise donc son essence en mécanique relativiste. Un test expérimental est proposé.

---

[1] W. Gerlach, O. Stern, "Der experimentelle Nachweis der Richtungsquantelung im Magnetfeld", Zeitschrift für Physik. 9 (1): 349–352, 1922.

[2] Le cours de physique, Electromagnétisme 2, Richard Feynman, Ed. Dunod Paris, p. 274, Chapitre 35, 1998.

[3] Catheline, S. A mechanistic interpretation of relativistic rigid body rotation. Sci Rep 13, 9047 (2023). https://doi.org/10.1038/s41598-023-35897-9.

[4] A. Papapetrou, "Spinning test-particles in general relativity. I," Proc. R. Soc. Lond. 209, 248 (1951)

[5] W. G. Dixon, "Dynamics of extended bodies in general relativity. I. Momentum and angular momentum," Proc. R. Soc. Lond. A 314, 499 (1970)

[6] J. Ehlers and E. Rudolph, "Dynamics of extended bodies in general relativity center-of-mass description and quasirigidity," Gen. Rel. Grav. 8, 197 (1977)

[7] L.Filipe, O.Costa, J.Natario, "Center of mass, spin supplementary conditions and the momentum of spinning particles", arXiv:1419.6443v2 (2015).

[8] R.C. Jennison, "Ray paths in a rotating system" Nature 199, 739 (1963).

[9] Ø. Grøn, "Relativistic description of a rotating disk with angular acceleration", Found. of Phys. 9, 353 (1979).

[10] Relativity in Rotating Frames, Eds. G. Rizzi and M.L. Ruggiero, Kluwer Academic Publishers (2005).

[11] Special Relativity in General Frames by E. Gourgoulhon, Ed. Springer 2010.

[12] Mathematiques Speciales, Electromagnétisme 4, M. Bertin, J.P.Faroux, J.Renault, Ed. Dunod Paris, p.154, Chap 7, 1984.

[13] Ø. Grøn, "Relativistic description of a rotating disk", Am. J. Phys. 43, 869 (1975).